\documentclass[11pt]{article}

\usepackage[utf8]{inputenc}
\usepackage{amssymb,amsmath,amsthm}
\usepackage{extarrows}
\usepackage{theoremref}
\usepackage{bm}

\newcount\hour \newcount\minutes \hour=\time \divide\hour by 60
\minutes=\hour \multiply\minutes by -60 \advance\minutes by \time
\def\mmmddyyyy{\ifcase\month\or Jan\or Feb\or Mar\or Apr\or May\or
Jun\or Jul\or Aug\or Sep\or Oct\or Nov\or Dec\fi \space\number\day,
\number\year}
\def\hhmm{\ifnum\hour<10 0\fi\number\hour :%
  \ifnum\minutes<10 0\fi\number\minutes} 
\makeatother

\newtheorem{definition}{Definition}
\newtheorem{theorem}[definition]{Theorem}
\newtheorem{corollary}[definition]{Corollary}
\newtheorem{lemma}[definition]{Lemma}

\newcommand{\Nat}{\mathbb{N}}
\newcommand{\Real}{\mathbb{R}}

\newcommand{\supp}{\mathrm{supp}}
\newcommand{\conv}{\mathrm{conv}}
\newcommand{\Sym}{\mathrm{Sym}}
\newcommand{\Orb}{\mathrm{Orb}}
\newcommand{\one}{\mathbf{1}}

\theoremstyle{remark}
\newtheorem{remark}{Remark}

\title{A Note on Polynomial Certificates for Walk Inequalities\footnote{Work done in part while the first author was at the University of Konstanz. Some of the material originates in the first author’s bachelor’s thesis \cite{willenborg-2016}.}}
\author{{\em Nadja Willenborg}\\
Institute of Computer Science \\
University of St.Gallen, Switzerland\\
{\tt Nadja.Willenborg@unisg.ch}
\and
{\em Sven Kosub} \\
Department of Computer and Information Science \\
University of Konstanz, Germany\\
{\tt Sven.Kosub@uni-konstanz.de}
}

\date{\empty}

\begin{document}

\maketitle


\begin{abstract}
Let \(w_m(G)\) denote the total number of walks of length \(m\) in an undirected graph \(G\). Spectral decomposition shows that  \((w_m(G))_{m\geq0}\) is a moment sequence of a finite positive measure. We use exchangeability of its product measures to turn global nonnegativity of polynomial symmetrizations into universal inequalities for the number of walks. For exponent vectors \(\alpha,\beta\) in distinct permutation orbits, \(\operatorname{Sym}(x^\beta-x^\alpha)\) is globally nonnegative exactly when \(\beta\) is coordinatewise even and majorizes \(\alpha\). 
This finite criterion includes several classical inequalities as special cases; univariate polynomials and squared alternants also yield linear and Hankel determinant inequalities.
\end{abstract}


\section{Introduction}

Let $\Nat_0=\{0,1,2,\dots\}$ be the set of nonnegative integers.
Let $G=(V,E)$ be an undirected graph on $n$ vertices with adjacency matrix $A=A(G)$.  For $m\in\Nat_0$, let
\[
  w_m(G)= \one^{\mathsf T}A^m\one 
\]
denote the total number of walks of length $m$ in $G$. Here, $\one$ is the $n$-dimensional all-ones vector. Equivalently, $w_m=w_m(G)$ is the number of graph homomorphisms from the path with $m$ edges to $G$.
Universal inequalities for the numbers $w_m$ therefore belong simultaneously to spectral graph theory and to the theory of graph homomorphism inequalities.

A basic example in this context is
\begin{equation}\label{eq:CS-intro}
  w_{a+b}^2\leq w_{2a}w_{2b},
  \qquad a,b\in\Nat_0,
\end{equation}
which follows from Cauchy--Schwarz (cf.~\cite{dress-gutman-2003}).  Broader families include the sandwich inequalities \cite{taeubig-et-al-2013}
\begin{equation}\label{eq:sandwich-intro}
  w_{2a+c}\,w_{2(a+b)+c}
  \leq w_{2a}\,w_{2(a+b+c)},
  \qquad a,b,c\in\Nat_0,
\end{equation}
and the generalized Erd\H{o}s--Simonovits inequalities
\begin{equation}\label{eq:general-ES-intro}
  w_{2a+b}^{k}
  \leq w_{2a+bk}\,w_{2a}^{k-1},
  \qquad k,a,b\in\Nat_0, k\ge 2, b\ge 1
\end{equation}
proved in~\cite{taeubig-et-al-2013} (see also
\cite{blakley-roy-1965,erdos-simonovits-1982,lagarias-et-al-1984} for precursor results).  Blekherman and Raymond subsequently described the cone of universal pure binomial inequalities for homomorphism counts from paths by means of tropicalization~\cite{blekherman-raymond-2022}.  
Moment and pseudomoment cones, sums of squares, and tropicalization have since been related in a general framework~\cite{blekherman-et-al-2025}.  
Here we specialize this framework to graph walk counts and derive, directly from symmetrization and majorization, an exact recognition criterion for symmetrized binomial certificates.

The note is organized around two results. 
First, building on the moment interpretation of walk
counts used systematically in \cite{barreras-et-al-2021}, Theorem 1 expresses every polynomial walk functional as an integral against an exchangeable
product measure. Second, Theorem \ref{thm:characterization} characterizes symmetrized binomial certificates by coordinatewise evenness and majorization, thereby giving a finite recognition test. Section~4 applies these results to classical and permutation inequalities, univariate linear certificates, and Hankel determinants, and exhibits a parity boundary for the certificate method.


\section{Spectral moments and the transfer principle}

Let the distinct eigenvalues of $A$ be denoted by $\lambda$, and let $E_\lambda$ be the corresponding orthogonal spectral projectors.  Define the finite positive measure
\begin{equation*}\label{eq:spectral-measure}
  \nu_G=_{\rm def}\sum_{\lambda\in\operatorname{spec}(A)}
        \lVert E_\lambda\one\rVert_2^2\,\delta_\lambda.
\end{equation*}
Its support consists precisely of the main eigenvalues of $G$, i.e., the eigenvalues whose eigenspaces are not orthogonal to $\one$.  Since $A^m=\sum_\lambda\lambda^mE_\lambda$, we have
\begin{equation}\label{eq:walk-moment}
  w_m(G)
  =\sum_\lambda \lambda^m\lVert E_\lambda\one\rVert_2^2
  =\int_{\Real}t^m\,d\nu_G(t).
\end{equation}
In particular, $\nu_G(\Real)=w_0(G)=|V|$.  The moment formula
\eqref{eq:walk-moment} is independent of choices inside repeated
eigenspaces.

For $r\geq 1$ and
\[
  f(x_1,\ldots,x_r)=\sum_{\gamma\in\Nat_0^r}c_\gamma
       x_1^{\gamma_1}\cdots x_r^{\gamma_r},
\]
where $c_\gamma \in \Real$ and $c_\gamma=0$ for all but finitely many $\gamma$, 
define the associated walk functional
\begin{equation*}\label{eq:walk-functional}
  \Phi_{G,r}(f)=_{\rm def}
  \sum_{\gamma\in\Nat_0^r}c_\gamma
       \prod_{i=1}^r w_{\gamma_i}(G).
\end{equation*}
We use normalized symmetrization
\begin{equation*}\label{eq:symmetrization}
  (\Sym_r f)(x_1,\ldots,x_r)
  =_{\rm def} \frac{1}{r!}\sum_{\sigma\in S_r}
       f(x_{\sigma(1)},\ldots,x_{\sigma(r)}),
\end{equation*}
where $S_r$ denotes the symmetric group on $r$ elements.

The following theorem constitutes the transfer principle.

\begin{theorem}\label{thm:transfer}
For every undirected graph $G$ and every polynomial
$f\in\Real[x_1,\ldots,x_r]$,
\begin{equation}\label{eq:transfer}
  \Phi_{G,r}(f)
  =\int_{\Real^r}f\,d\nu_G^{\otimes r}
  =\int_{\Real^r}\Sym_r f\,d\nu_G^{\otimes r}
  =\Phi_{G,r}(\Sym_r f).
\end{equation}
Consequently, if $\Sym_r f$ is globally nonnegative, then
$\Phi_{G,r}(f)\geq 0$ for every $G$.
\end{theorem}

\begin{proof}
Using (\ref{eq:walk-moment}) and the product structure of
$\nu_G^{\otimes r}$, we obtain
\[
  \int_{\Real^r}f\,d\nu_G^{\otimes r}
  =\sum_\gamma c_\gamma
       \prod_{i=1}^r\int_\Real x_i^{\gamma_i}\,d\nu_G(x_i)
  =\Phi_{G,r}(f).
\]
As the product measure is invariant under every coordinate permutation, averaging the integral over $S_r$ gives the second equality in (\ref{eq:transfer}). The final assertion follows from positivity of the measure.
\end{proof}

Let $D(G)$ denote the difference between the right- and
left-hand sides of a prescribed walk inequality for graph $G$. A globally
nonnegative polynomial $p\in\mathbb R[x_1,\ldots,x_r]$ is called
a polynomial certificate for $D(G)\geq0$ if and only if 
\[
  D(G)=\Phi_{G,r}(p)
\]
for every undirected graph $G$. 
Note that no SOS representation of $p$ is required.

\begin{remark}\label{rem:equality-transfer}
If $\Sym_r f\geq 0$, then equality in the certified walk inequality holds
for $G$ if and only if $\Sym_r f$ vanishes on
$\supp(\nu_G)^r$.  This follows because $\nu_G$ is an atomic measure with
strictly positive mass at every point of its support.
\end{remark}

\begin{remark}
Nothing in Theorem \ref{thm:transfer} is specific to $0$--$1$ adjacency matrices. The same statement holds for the moment sequence $s_m=u^{\mathsf T}B^mu$ of any real symmetric matrix $B$ and vector $u$, with spectral measure $\sum_\lambda\lVert E_\lambda u\rVert_2^2\delta_\lambda$. The graph case is obtained from $B=A(G)$ and $u=\one$.
\end{remark}


\section{An exact majorization criterion for binomial certificates}

For $\gamma\in\Nat_0^r$, let
\[|\gamma|=_{\rm def} \sum_{i=1}^r \gamma_i, 
   \qquad
  x^\gamma=_{\rm def}\prod_{i=1}^r x_i^{\gamma_i},
  \qquad
  W_\gamma(G)=_{\rm def}\prod_{i=1}^r w_{\gamma_i}(G),
\]
and define the normalized symmetric monomial
\begin{equation*}\label{eq:monomial-mean}
  m_\gamma(x)=_{\rm def} \Sym_r(x^\gamma)
  ~~=\frac{1}{r!}\sum_{\sigma\in S_r}
      \prod_{i=1}^r x_i^{\gamma_{\sigma(i)}}.
\end{equation*}
This normalization has the useful property
\begin{equation}\label{eq:Phi-monomial}
  \Phi_{G,r}(m_\gamma)=W_\gamma(G).
\end{equation}

We write $\gamma^\downarrow$ for the coordinates of $\gamma$ in
nonincreasing order.  
For $\alpha,\beta\in\Real^r$, the vector $\beta$ {\em majorizes} $\alpha$, in symbols $\beta\succeq\alpha$, if and only if 
\begin{equation}\label{eq:majorization}
  \sum_{i=1}^j\beta_i^\downarrow
  \geq\sum_{i=1}^j\alpha_i^\downarrow\quad
  \textrm{for all $1\le j<r$},
  \qquad
  \sum_{i=1}^r\beta_i=\sum_{i=1}^r\alpha_i.
\end{equation}
We use the following standard facts from majorization theory (see \cite[Sections~2.B, 3.G, and 4.C]{marshall-olkin-arnold-2011}). For $\alpha,\beta\in\Real^r$,
\[
  \beta\succeq\alpha
  ~\Longleftrightarrow~
  \alpha=D\beta
  \text{ for some doubly stochastic matrix }D
  ~\Longleftrightarrow~
  \alpha\in\conv\ \Orb(\beta),
\]
where $\Orb(\beta)=_{\rm def}\{~(\beta_{\sigma(1)}, \dots, \beta_{\sigma(r)})~|~\sigma\in S_r\}$ denotes the permutation orbit of $\beta$.
Moreover, for $\alpha,\beta\in\Nat_0^r$ of equal total degree,
\begin{equation}
  \beta\succeq\alpha
  ~\Longleftrightarrow~
  m_\beta(x)\geq m_\alpha(x)
  \quad\text{for all }x\in\Real_{>0}^r. \label{eq:muirhead}
\end{equation}
The first line combines the Hardy--Littlewood--Pólya and Rado
characterizations of majorization; the implication from left to right in the second line is Muirhead's inequality.  The reverse implication follows directly by strict separation: if $\alpha\notin\conv\ \Orb(\beta)$, choose $u\in\Real^r$ separating $\alpha$ from $\conv\ \Orb(\beta)$, i.e., $u^{\mathsf T}\alpha> \max_{\delta\in\operatorname{Orb}(\beta)}u^{\mathsf T}\delta$, substitute $x_i=t^{u_i}$, and let $t\to\infty$.

We also need a standard elementary fact about globally nonnegative polynomials.  

\begin{lemma}\label{lem:vertex-terms}
Let $p(x)=\sum_\gamma c_\gamma x^\gamma$ be globally nonnegative and not identically zero.  If $\eta$ is a vertex of the Newton polytope
$\conv\ \{~\gamma~|~c_\gamma\neq 0~\}$, then $c_\eta>0$ and every coordinate of
$\eta$ is even.
\end{lemma}

\begin{proof}
Choose $u\in\Real^r$ exposing $\eta$, so that
$u^{\mathsf T}\eta>u^{\mathsf T}\gamma$ for every other exponent $\gamma$ in the
support.  For a sign vector $s\in\{-1,1\}^r$, substitute
$x_i=s_i t^{u_i}$ and divide by $t^{u\cdot\eta}$.  As $t\to\infty$, the
result tends to $c_\eta s^\eta$.  Nonnegativity of $p$ forces this limit to
be nonnegative for every $s$.  Since $c_\eta\neq 0$, this is possible only
when $\eta$ is coordinatewise even and $c_\eta>0$.
\end{proof}

For $\alpha,\beta\in\mathbb N_0^r$, we call
\[
 C_{\alpha,\beta}
 =_{\rm def} m_\beta-m_\alpha
 =\operatorname{Sym}_r(x^\beta-x^\alpha)
\]
the symmetrized binomial associated with the exponent
presentation $(\alpha,\beta)$. It is a polynomial certificate for
$W_\alpha\leq W_\beta$ precisely when it is globally nonnegative.

The next theorem characterizes when $C_{\alpha,\beta}$ is globally nonnegative.
The exclusion of equal permutation orbits is essential: if $\alpha$ is a permutation of $\beta$, then $m_\beta-m_\alpha=0$ regardless of parity.

\begin{theorem}
\label{thm:characterization}
Let $\alpha,\beta\in\Nat_0^r$ and assume that $\alpha$ is not a coordinate permutation of $\beta$.  Then, 
\[
m_\beta-m_\alpha \textrm{ is globally nonnegative on $\Real^r$}
~\Longleftrightarrow~ 
\beta\in(2\Nat_0)^r \textrm{ and $\beta\succeq\alpha$}.
\]
\end{theorem}

\begin{proof}
For $(\Rightarrow)$, assume first that $p=_{\rm def} m_\beta-m_\alpha$ is globally nonnegative. Evaluation at $(t,\ldots,t)$ gives
$p(t,\ldots,t)=t^{|\beta|}-t^{|\alpha|}\geq 0$ for all $t>0$, 
hence $|\alpha|=|\beta|$.  Since $p$ is nonnegative on
$\Real_{>0}^r$, the reverse implication in \eqref{eq:muirhead}
immediately yields $\beta\succeq\alpha$.
It remains to prove that $\beta$ is coordinatewise even.
Let $P_\beta=\conv\ \Orb(\beta)$.  By the majorization already
established and Rado's theorem, $\Orb(\alpha)\subseteq P_\beta$. Hence the Newton polytope of $p$ is $P_\beta$.  Every distinct point of $\Orb(\beta)$ is a vertex of $P_\beta$: all orbit points have the same Euclidean norm, and for
$\gamma\in\Orb(\beta)$ the functional $z\mapsto\gamma^{\mathsf T}z$ is uniquely maximized over the distinct orbit points at $\gamma$.  Each such
point has a positive coefficient in $p$.  By Lemma
\ref{lem:vertex-terms}, every such point is coordinatewise even.  In
particular, $\beta\in(2\Nat_0)^r$.

For $(\Leftarrow)$, suppose that $\beta$ is coordinatewise even and $\beta\succeq\alpha$.  For arbitrary $x\in\Real^r$, write
$|x|=(|x_1|,\ldots,|x_r|)$. We obtain $m_\alpha(x)\leq m_\alpha(|x|)$ and $m_\beta(x)=m_\beta(|x|)$.
Muirhead's inequality, extended from the positive to the nonnegative orthant by continuity, now yields
\[
  m_\alpha(x)
  \leq m_\alpha(|x|)
  \leq m_\beta(|x|)
  =m_\beta(x).
\]
Thus $m_\beta-m_\alpha$ is globally nonnegative.
\end{proof}

\begin{remark}\label{rem:general-moments}
After cancelling common moment factors (cf. Remark 5), the evenness-and-majorization criterion coincides with the univariate balanced specialization—equal numbers of moment factors and equal total index—of the classification of pure binomial inequalities in absolute moments in [3, Corollary 4.4, Theorem 4.12, and Lemmas 4.13–4.14]. Thus, at the level of universal moment inequalities, the criterion is already contained in [3]. Theorem 3 provides a pointwise symmetric polynomial realization of the same criterion.
\end{remark}

Combining Theorem \ref{thm:characterization} with the transfer principle gives the following majorization family of walk inequalities.

\begin{corollary}\label{cor:majorization-walks}
If $\alpha,\beta\in\Nat_0^r$, $\beta$ is coordinatewise even, and
$\beta\succeq\alpha$, then
\begin{equation*}\label{eq:majorization-walks}
  W_\alpha(G)=\prod_{i=1}^r w_{\alpha_i}(G)
  \leq
  \prod_{i=1}^r w_{\beta_i}(G)=W_\beta(G)
\end{equation*}
for every undirected graph $G$.
\end{corollary}
\begin{proof}
The case of equal orbits is trivial. Otherwise, 
apply Theorem \ref{thm:characterization} and Theorem \ref{thm:transfer} to $m_\beta-m_\alpha$ and use \eqref{eq:Phi-monomial}.
\end{proof}

\begin{remark}
For two distinct exponent orbits, certifiability by the present method is decided by checking that every coordinate of $\beta$ is even, sorting $\alpha$ and $\beta$, and comparing $r$ partial sums in \eqref{eq:majorization}. 
Thus the original real-algebraic question reduces to a finite combinatorial test.
\end{remark}

\begin{remark}\label{rem:common-factors}
The symmetrized binomial depends on the chosen factor tuple.  Before applying the recognition test, it is natural to delete the multiset intersection of $\{\alpha_1,\ldots,\alpha_r\}$ and
$\{\beta_1,\ldots,\beta_r\}$.  
Indeed, multiplying an already certified inequality by a common nonnegative walk count preserves validity, although
the symmetrized binomial in the enlarged set of variables need not remain globally nonnegative.  For example, $w_0w_4\geq w_2^2$ has a two-variable certificate, so $w_0w_4w_1\geq w_2^2w_1$ is valid; the unreduced exponent tuple on the majorizing side is $(0,4,1)$ and is nevertheless not even.
The reduced convention also matches the disjoint-support convention for pure binomial moment inequalities in~\cite{blekherman-et-al-2025}.
\end{remark}


\section{Consequences}

\subsection{Extremal and two-factor inequalities}

The vector $(2d,0,\ldots,0)$ majorizes every nonnegative vector with total sum $2d$.  Therefore Corollary \ref{cor:majorization-walks} immediately gives a weighted arithmetic--geometric-mean type inequality.

\begin{corollary}\label{cor:AG-walks}
Let $\alpha\in\Nat_0^r$ with $|\alpha|=2d$, $d\in\Nat_0$.  Then, for each undirected graph $G$,
\begin{equation*}\label{eq:AG-walks}
  \prod_{i=1}^r w_{\alpha_i}
  \leq w_0^{r-1}w_{2d}.
\end{equation*}
\end{corollary}

For $r=2$, the characterization becomes especially explicit and recovers exactly the nontrivial members of the sandwich family in \eqref{eq:sandwich-intro}.

\begin{corollary}\label{cor:two-factor}
Let $\alpha,\beta\in\Nat_0^2$ belong to different permutation orbits. After independently permuting the coordinates of $\alpha$ and $\beta$, $m_\beta-m_\alpha$ is globally nonnegative if and only if there are $a,b,c\in\Nat_0$ with $c>0$ such that
\begin{equation*}\label{eq:two-factor-parameters}
    \beta=(2a,\,2a+2b+2c), \qquad
    \alpha=(2a+c,\,2a+2b+c).
\end{equation*}
Consequently,
\[
  w_{2a+c}\,w_{2(a+b)+c}
  \leq w_{2a}\,w_{2(a+b+c)}.
\]
\end{corollary}

\begin{proof}
Order both vectors increasingly.  The conditions in
Theorem \ref{thm:characterization} say precisely that the endpoints $\beta_1$ and $\beta_2$ are even, that
$\beta_1\leq\alpha_1\leq\alpha_2\leq\beta_2$, and that the two vectors have the same sum.  Set
\[
  a=\frac{\beta_1}{2},\qquad
  c=\alpha_1-\beta_1,\qquad
  b=\frac{\alpha_2-\alpha_1}{2}.
\]
The parity of $\alpha_1$ and $\alpha_2$ is equal because their sum is even, so $a,b,c\in\Nat_0$.  Distinctness of the orbits gives $c>0$, and the last coordinate of $\beta$ follows from equality of the sums.

For completeness, the corresponding polynomial factors as
\begin{equation*}\label{eq:two-factor-factorization}
  2(m_\beta-m_\alpha)(x,y)
  =x^{2a}y^{2a}
    \left(x^{2b+c}-y^{2b+c}\right)(x^c-y^c).
\end{equation*}
The last two factors have the same sign: for odd $c$ both powers are odd and increasing; for even $c$ both are increasing functions of the absolute value.  This also proves sufficiency directly.
\end{proof}

As a three-factor illustration, $(8,2,0)\succeq(6,3,1)$ and the first vector is even.  Hence every graph satisfies
\begin{equation}\label{eq:three-factor-example}
  w_6w_3w_1\leq w_8w_2w_0.
\end{equation}
This example is outside the special permutation construction in the next
subsection: for $\beta/2=(4,1,0)$, no vector
$\beta/2+P_\sigma(\beta/2)$ is a permutation of $(6,3,1)$.  The point is not the isolated inequality, but that all such examples are generated and recognized by the same partial-sum criterion.

\subsection{Permutation inequalities}

Let $\alpha\in\Nat_0^r$ and $\sigma\in S_r$, and set
$\gamma_i=_{\rm def}\alpha_i+\alpha_{\sigma(i)}$.  If $P_\sigma$ is the permutation matrix acting by $(P_\sigma\alpha)_i=\alpha_{\sigma(i)}$, then
\[
  \gamma=\frac{I+P_\sigma}{2}(2\alpha).
\]
The matrix $(I+P_\sigma)/2$ is doubly stochastic.  Thus
$2\alpha\succeq\gamma$, and Corollary \ref{cor:majorization-walks} yields the next statement.

\begin{corollary}\label{cor:permutation}
Let $\alpha\in\Nat_0^r$ and $\sigma\in S_r$. Then, for every undirected graph $G$,
\begin{equation}\label{eq:permutation}
  \prod_{i=1}^r w_{\alpha_i+\alpha_{\sigma(i)}}
  \leq \prod_{i=1}^r w_{2\alpha_i}.
\end{equation}
Moreover, the gap has the exact single-square representation
\begin{equation}\label{eq:permutation-gap}
  \prod_{i=1}^r w_{2\alpha_i}
  -\prod_{i=1}^r w_{\alpha_i+\alpha_{\sigma(i)}}
  =\frac12\int_{\Real^r}
    \left(
      \prod_{i=1}^r x_i^{\alpha_i}
      -\prod_{i=1}^r x_i^{\alpha_{\sigma(i)}}
    \right)^2
    d\nu_G^{\otimes r}(x).
\end{equation}
\end{corollary}

\begin{proof}
Only (\ref{eq:permutation-gap}) remains to be shown.  Expanding the square and integrating term by term gives twice the right-hand product in (\ref{eq:permutation}) minus twice its left-hand product, because $\sigma$ merely permutes the factors in the second square term.
\end{proof}

The square on the right-hand side of \eqref{eq:permutation-gap} provides a compact single-square certificate and an exact equality test via Remark \ref{rem:equality-transfer}.
For example, a $d$-regular graph satisfies $\nu_G=|V|\delta_d$ because $A\one=d\one$, and therefore attains equality in (\ref{eq:permutation}).

Note that \eqref{eq:permutation} also follows by applying \eqref{eq:CS-intro} to each pair $(\alpha_i,\alpha_{\sigma(i)})$. We obtain
\[
  w_{\alpha_i+\alpha_{\sigma(i)}}^2
  \leq w_{2\alpha_i}w_{2\alpha_{\sigma(i)}}.
\]
Multiplying over $i$ and using permutation invariance of the
factors on the right yields \eqref{eq:permutation} after taking square roots.

\subsection{Beyond binomials: univariate and determinantal certificates}

The linear and determinantal inequalities below are two manifestations of the same Hankel matrix positivity (cf.~\cite{barreras-et-al-2021}). 

\paragraph{Linear inequalities.} For $k\geq 0$, set
\[
  v_k(x)=_{\rm def} (1,x,\ldots,x^k)^{\mathsf T}
  \quad\text{and}\quad
  H_k(G)=_{\rm def}\bigl(w_{i+j}(G)\bigr)_{i,j=0}^k.
\]
The moment representation gives
\begin{equation*}\label{eq:hankel-gram}
  H_k(G)
  =\int_{\Real}v_k(x)v_k(x)^{\mathsf T}\,d\nu_G(x)
  \succeq 0.
\end{equation*}
Every globally nonnegative univariate polynomial $p$ of degree at most
$2k$ is a sum of squares of real polynomials of degree at most $k$.
Equivalently,
\[
  p(x)=v_k(x)^{\mathsf T}Qv_k(x)
\]
for some, generally nonunique, positive semidefinite matrix $Q$.  Therefore
\begin{equation*}\label{eq:linear-hankel-pairing}
  \Phi_{G,1}(p)
  =\int_{\Real}p(x)\,d\nu_G(x)
  =\operatorname{tr}\bigl(QH_k(G)\bigr)
  \geq 0.
\end{equation*}
Conversely, requiring $\Phi_{G,1}(q^2) \ge 0$ for all polynomials $q$ of degree at most $k$ is exactly the condition $H_k(G)\succeq 0$.

The following family makes this connection explicit.
Let $d=2k$ with $k\geq 1$, and let
\begin{equation}\label{eq:univariate-polynomial}
  f(x)=x^d-\sum_{j=1}^{d-1}a_jx^{d-j},
  \qquad \textrm{$a_j\geq 0$},
\end{equation}
where not all $a_j$ vanish.  The derivative $f'$ has a unique positive zero $\gamma$.  Indeed, after zero coefficients are omitted, the coefficients of $f'$ have exactly one sign change, while $f'$ is negative for sufficiently small positive $x$ and positive for sufficiently large $x$.  Descartes' rule of signs therefore gives the assertion, and $\gamma$ is the unique minimizer of $f$ on $[0,\infty)$.
Since $d$ is even,
\[
  f(-x)-f(x)
  =2\sum_{\substack{1\leq j\leq d-1\\ j\ \mathrm{odd}}}
       a_jx^{d-j}
  \geq 0, 
  \qquad x\geq 0,
\]
so $\gamma$ is also a global minimizer on $\Real$.  Set
\begin{equation*}\label{eq:univariate-constant}
  c=_{\rm def}-f(\gamma)
    ~~=\frac{1}{d}\sum_{j=1}^{d-1}j\,a_j\gamma^{d-j}>0,
\end{equation*}
where the second identity follows from $f'(\gamma)=0$.  The polynomial $f+c$ is globally nonnegative, and hence
\begin{equation*}\label{eq:univariate-family}
  w_d+c\,w_0
  \geq \sum_{j=1}^{d-1}a_jw_{d-j}
\end{equation*}
for every undirected graph.  Equivalently, $f+c=v_k^{\mathsf T}Qv_k$ for some $Q\succeq0$, and the difference between the two sides of the inequality is exactly 
\[
\operatorname{tr}(QH_k(G))
=\sum_{r,s=0}^kQ_{rs}w_{r+s}
=w_d+cw_0-\sum_{j=1}^{d-1}a_jw_{d-j} \ge 0.
\]

\medskip

\paragraph{Determinantal inequalities.} 
For $s\ge 1$, let $q_1<\ldots<q_s$ be nonnegative integers. Define
\[
  v_q(x)=_{\rm def}(x^{q_1},\dots,x^{q_s})^{\mathsf T}
\]
and the generalized Hankel moment matrix
\begin{equation*}\label{eq:moment-matrix}
  H_q(G)
  =_{\rm def}\bigl(w_{q_i+q_j}(G)\bigr)_{i,j=1}^s
  ~~=\int_{\Real}v_q(x)v_q(x)^{\mathsf T}\,d\nu_G(x)
  \succeq0.
\end{equation*}
Define $A_q(x_1,\ldots,x_s)=_{\rm def} \det\bigl(x_j^{q_i}\bigr)_{i,j=1}^s$.
Expanding \(A_q^2\) and integrating term by term gives
\begin{equation*}\label{eq:andreief}
  \det H_q(G)
  =\frac{1}{s!}\Phi_{G,s}(A_q^2)
  =\frac{1}{s!}\int_{\Real^s}
     \left[\det\bigl(x_j^{q_i}\bigr)_{i,j=1}^s\right]^2
     d\nu_G^{\otimes s}(x)
  \geq0.
\end{equation*}
For $q_i=i-1$, the alternant is the Vandermonde determinant, and hence
\begin{equation}\label{eq:hankel-determinant}
  \det\bigl(w_{i+j}\bigr)_{i,j=0}^{s-1}
  =\frac{1}{s!}\int_{\Real^s}
     \prod_{1\leq i<j\leq s}(x_j-x_i)^2
     d\nu_G^{\otimes s}(x)
  \geq0.
\end{equation}
For $s=3$, this becomes
\begin{equation*}\label{eq:hankel-3}
  w_0w_2w_4+2w_1w_2w_3
  \geq w_0w_3^2+w_1^2w_4+w_2^3.
\end{equation*}
In the Vandermonde case, the determinant in
\eqref{eq:hankel-determinant} is positive exactly when $G$ has at least $s$ distinct main eigenvalues: the integrand is positive precisely on tuples of pairwise distinct points in the support of $\nu_G$.

\subsection{A parity obstruction for symmetrized binomial certificates}

Theorem \ref{thm:characterization} characterizes one specific proof mechanism: global nonnegativity of the symmetrized binomial $m_\beta-m_\alpha$.  It does not characterize all inequalities valid on walk sequences of simple graphs.  Spectral measures arising from graphs
form a restricted subclass of all positive measures, and valid inequalities may also follow by composing several certificates.

The generalized Erd\H{o}s--Simonovits family illustrates this distinction.  For $k, a,b\in\Nat_0$, $k \ge 2$ and $b\geq 1$, write
\[
  \alpha=(2a+b,\ldots,2a+b),
  \qquad
  \beta=(2a+bk,2a,\ldots,2a).
\]
Then $W_\alpha\leq W_\beta$ is precisely \eqref{eq:general-ES-intro}, and $\beta\succeq\alpha$ for every choice of
parameters.  The parity part of Theorem \ref{thm:characterization} is therefore the only remaining issue.

\begin{corollary}\label{cor:ES-boundary}
For $k, a,b\in\Nat_0$, $k \ge 2$ and $b\geq 1$, 
the symmetrized binomial for (3) is globally
nonnegative if and only if $bk$ is even.
In particular, when both $k$ and $b$ are odd, the inequality is
universally valid by~{\em \cite{taeubig-et-al-2013}} but has no certificate of this form.
\end{corollary}

\begin{proof}
Majorization is automatic, the two exponent vectors belong to distinct permutation orbits, and $\beta$ is coordinatewise even exactly when $2a+bk$, equivalently $bk$, is even.
\end{proof}

\section{Conclusion}


The main contribution of this note is the pointwise symmetric polynomial formulation of the evenness-and-majorization criterion and its transfer to walk inequalities.
Applications recover or repackage standard consequences of Muirhead’s inequality, Cauchy–Schwarz, and Hankel matrix positivity within this common framework.
However, the criterion characterizes only globally nonnegative symmetrized binomial certificates, not all universally valid walk inequalities. Corollary 8 exhibits this limitation: for odd $b$ and $k$, the generalized Erdős–Simonovits inequality remains valid although its natural symmetrized polynomial is not globally nonnegative. A concrete open problem is therefore to characterize, for fixed $r$ and $d$, the symmetric polynomials $p$ of degree at most $d$ satisfying
\[
\Phi_{G,r}(p)\geq0
\]
for every undirected graph $G$, and to develop graph-specific certificates that exploit the entrywise nonnegativity of adjacency matrices.

\paragraph{Acknowledgment.}
We thank Markus Schweighofer and Tobias T\"opfer for helpful hints.

\end{document}